\documentclass[preprint,12pt]{elsarticle}

\usepackage{amssymb}
\usepackage{amsmath}

\journal{ }

\begin{document}

\begin{frontmatter}



\title{Rational solutions of (1+1)-dimensional Burgers equation and their asymptotic}

	\author[label1]{V.I. Avrutskiy}
\ead{avrutsky@phystech.edu}
\author[label1]{V.P. Krainov}
\ead{vpkrainov@mail.ru}
\address[label1]{Moscow Institute of Physics and Technology}

\begin{abstract}
A special initial condition for (1+1)-dimensional Burgers equation
is considered. It allows to obtain new analytical solutions for an arbitrary
low viscosity as well as for the inviscid case. The viscous solution
is written as a rational function provided the Reynolds number (a
dimensionless value inversely proportional to the viscosity) is a multiple
of two. The inviscid solution is expressed in radicals. Asymptotic expansion
of the viscous solution at infinite Reynolds number is compared against
the inviscid case. All solutions are finite, tend to zero at infinity
and therefore are physically viable.
\end{abstract}
\begin{keyword}
analytical solutions \sep nonlinear differential equations \sep Burgers equation \sep asymptotic expansion


\end{keyword}

\end{frontmatter}


\section{Introduction}
Initially written to simulate turbulence \cite{rudenko2016self}, Burgers equation \cite{burgers2013nonlinear} was later used to describe many nonlinear dissipative processes without dispersion, including formation of the large scale structure of the Universe \cite{Gurbatov:2012}. Its (1+1)-dimensional form has numerous analytical solutions \cite{benton1972table,dhawan2012contemporary,kudryashov2010comment,rodin1970riccati,rodin1970some,rudenko2016self,samokhin2016burgers,wazwaz2007variational,wood2006exact}.
It can be viewed as the simplest analogue of the Navier-Stokes equations, and therefore is widely used for benchmarking numerical methods \cite{guzhev1995burgers}. For this purpose, a reference solution should develop arbitrarily large gradients, a proper capturing of which is rigorously tested. A popular reference solution was obtained by Cole \cite{cole1951quasi} in a form of a series which has some convergence problems. Alternatively it can be expressed as an integral which has to be numerically calculated in each point of the solving domain \cite{basdevant1986spectral}. In this letter propose an analytical solution in a form of a rational expression. This solution can develop arbitrarily large gradients.
\section{The equation}
Consider an initial value problem for (1+1)-dimensional Burgers equation
\begin{equation}
\frac{\partial v}{\partial t}+v\frac{\partial v}{\partial x}=\nu\frac{\partial^{2}v}{\partial x^{2}}\label{eq:burgers}
\end{equation}
solved for $z\in[-\infty,\infty]$, $t\in[0,\infty]$ with initial
condition
\begin{equation}
v(0,x)=-\frac{2v_{0}(x/a)}{1+(x/a)^{2}},\label{eq:initc}
\end{equation}
here $v_{0}$ and $a$ are arbitrary positive constants. Using dimensionless
variables
\[
v=v_{0}V,\quad t=\frac{a}{v_{0}}T,\quad x=aZ,
\]
the equation and initial condition can be written as
\begin{equation}
\frac{\partial V}{\partial T}+V\frac{\partial V}{\partial Z}=\frac{1}{s}\frac{\partial^{2}V}{\partial Z^{2}},\label{eq:bezr}
\end{equation}
\begin{equation}\label{eq:init}
V(0,Z)=-\frac{2Z}{1+Z^{2}}.
\end{equation}
Here $s=v_{0}a/\nu$ is a dimensionless parameter similar to the Reynolds
number. In Section 3 this equation will be solved for cases when $s$
is a multiple of two. In Section 4 it will be solved without the viscous
term ($s=\infty$). In Section 5 we examine the asymptotic expansions of
the spacial gradient $\partial V/\partial Z$ of this solution at $s\rightarrow\infty$, $Z=0$ and
also write down the asymptotic of the maximum gradient (by absolute value) of the viscous solution.
\begin{figure}[h] \centering
	\includegraphics[height=0.3\textheight]{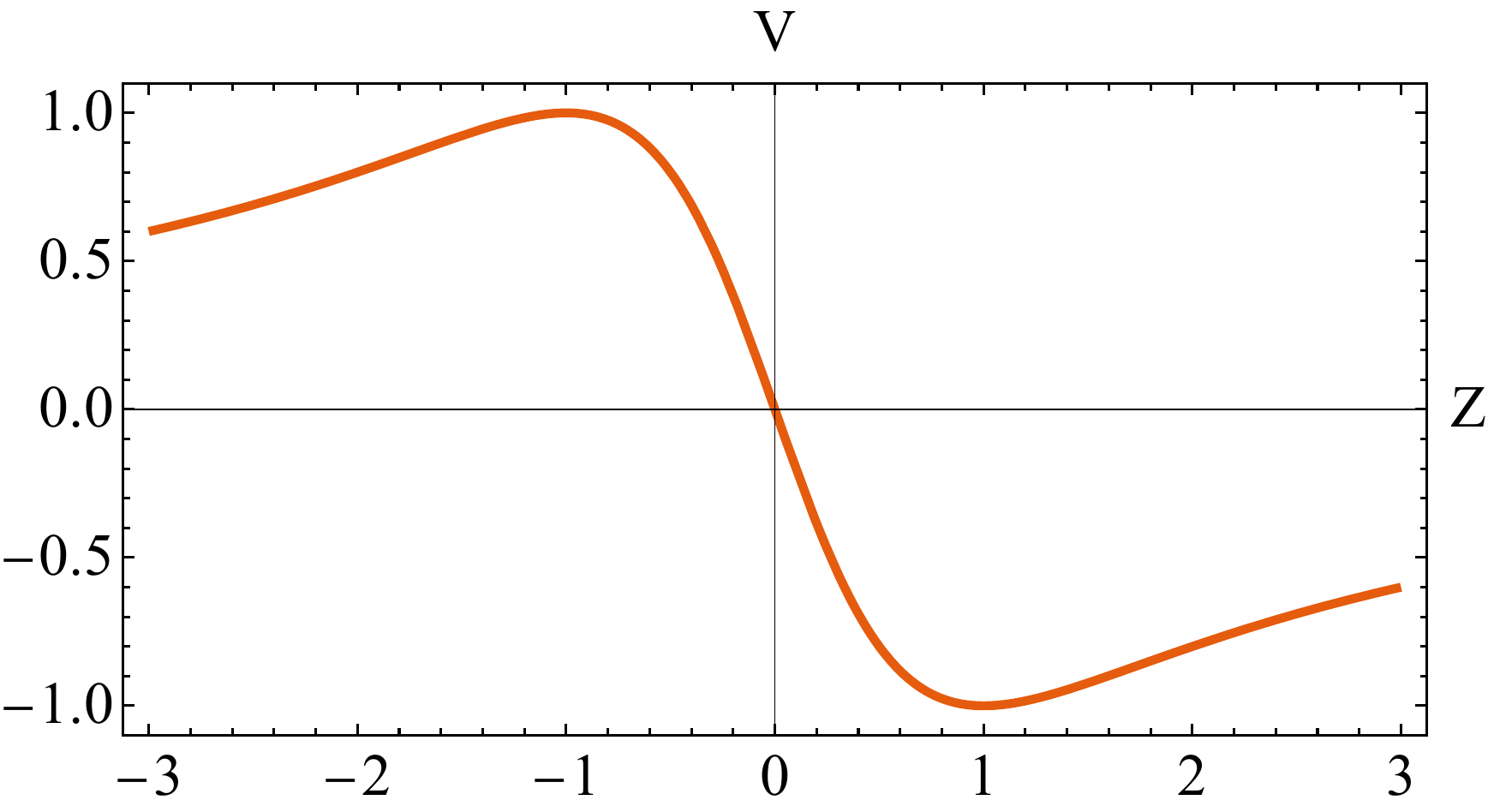}
	\caption{Initial condition}
\end{figure}
\section{Case $s/2\in\mathbb{Z}$}
The Hopf-Cole transformation \cite{hopf1950partial,cole1951quasi}
\begin{equation}
V=-\frac{2}{s}\frac{\partial}{\partial Z}\ln W\label{eq:zamena}
\end{equation}
applied to (\ref{eq:bezr}) yields a diffusion equation
\begin{equation}
\frac{\partial W}{\partial T}=\frac{1}{s}\frac{\partial^{2}W}{\partial Z^{2}}.\label{eq:diffusion}
\end{equation}
The particular form (\ref{eq:init}) of the initial condition for $V$ allows it to
escape the exponent in the initial condition for $W$
\begin{equation}
W(0,Z)=C\exp\left[-\frac{s}{2}\int V(0,Z)dZ\right]=C\exp\left[\frac{s}{2}\ln(1+Z^{2})\right]=C(1+Z^{2})^{s/2}.\label{eq:initw}
\end{equation}
The solution of (\ref{eq:diffusion}) can be written in a standard form (for brevity $r=s/2$)
\begin{equation}\label{integral}
W(T,Z)=C\sqrt{\frac{r}{2\pi T}}\int_{-\infty}^{\infty}\exp[-\frac{r}{2T}(Z-\widehat{Z})^{2}](1+\widehat{Z}^{2})^{r}d\widehat{Z}.
\end{equation}
If $r$ is integer the expression $(1+\widehat{Z}^{2})^{r}$ can be
expanded using binomial coefficient
\[
(1+\widehat{Z}^{2})^{r}=\sum_{\alpha=0}^{r}\binom{r}{\alpha}\widehat{Z}^{2\alpha},
\]
and the solution $W(T,Z)$ breaks into a sum of integrals $I_{\alpha}$
(for breivity $\beta=r/T$)
\[
W=\frac{C\sqrt{\beta}}{\sqrt{2\pi}}\sum_{\alpha=0}^{r}\binom{r}{\alpha}I_{\alpha}
\]
\[
I_{\alpha}=\int_{-\infty}^{\infty}\exp[-\frac{\beta}{2}(Z-\widehat{Z})^{2}]\widehat{Z}^{2\alpha}d\widehat{Z}.
\]
Each integral is a polynomial in $Z^{2}$
\[
I_{\alpha}=\frac{\sqrt{2\pi}}{\beta^{\alpha+1/2}}\frac{(2\alpha)!}{2^{\alpha}\alpha!}\sum_{k=0}^{\alpha}(-\beta Z^{2})^{k}\frac{(-\alpha)_{k}}{2^{k}k!(1/2)_{k}},
\]
where $(a)_{b}\equiv a(a+1)(a+2)\ldots(a+b-1)$ is the Pochhammer
symbol. Thus, $W$ can be written using double summation
\begin{equation}
W=C\sum_{\alpha=0}^{r}\left[\binom{r}{\alpha}\frac{(2\alpha)!}{2^{\alpha}\alpha!}\frac{1}{\beta^{\alpha}}\sum_{k=0}^{\alpha}(-\beta Z^{2})^{k}\frac{(-\alpha)_{k}}{2^{k}k!(1/2)_{k}}\right].\label{eq:summa}
\end{equation}
It is useful to separate $Z$ and $\beta$ by changing the order of summation
\[
W=C\sum_{k=0}^{r}\left[\frac{(-1)^{k}Z^{2k}}{2^{k}k!(1/2)_{k}}\sum_{\alpha=k}^{r}\binom{r}{\alpha}\frac{(-\alpha)_{k}(2\alpha)!}{2^{\alpha}\alpha!}\left(\frac{1}{\beta}\right)^{\alpha-k}\right].
\]
After substituting $\beta=r/T$, $r=s/2$ and simplifying the expression we obtain
\begin{equation}
W=C\sum_{k=0}^{s/2}\left[\frac{Z^{2k}}{2^{k}k!(2k-1)!!}\sum_{\alpha=k}^{s/2}\frac{(2\alpha)!}{\alpha!(s/2-\alpha)!(\alpha-k)!}\left(\frac{T}{s}\right)^{\alpha-k}\right].\label{eq:W}
\end{equation}
Thus, $W$ is a polynomial of order $s/2$ in $Z^{2}$ and the coefficient
of $(Z^{2})^{k}$ is a polynomial of order $s/2-k$ in $T$. According to (\ref{eq:zamena}), the relation between $V$ and $W$ is 
\begin{equation}
V=-\frac{2}{s}\frac{\partial W/\partial Z}{W}.\label{eq:svyaz}
\end{equation}
Therefore, the solution of (\ref{eq:bezr}) for even $s$ is
\begin{equation}
V(s,T,Z)=-\frac{2}{s}\frac{\sum_{k=0}^{s/2}\left[\frac{Z^{2k-1}}{2^{k-1}(k-1)!(2k-1)!!}\sum_{\alpha=k}^{s/2}\frac{(2\alpha)!}{\alpha!(s/2-\alpha)!(\alpha-k)!}\left(\frac{T}{s}\right)^{\alpha-k}\right]}{\sum_{k=0}^{s/2}\left[\frac{Z^{2k}}{2^{k}k!(2k-1)!!}\sum_{\alpha=k}^{s/2}\frac{(2\alpha)!}{\alpha!(s/2-\alpha)!(\alpha-k)!}\left(\frac{T}{s}\right)^{\alpha-k}\right]}\label{eq:otvet}
\end{equation}
It is worth noticing that for rational $T,Z$ this expression is also
a rational number which can be useful for numerical applications.
\begin{figure}[h] \centering
	\includegraphics[height=0.35\textheight]{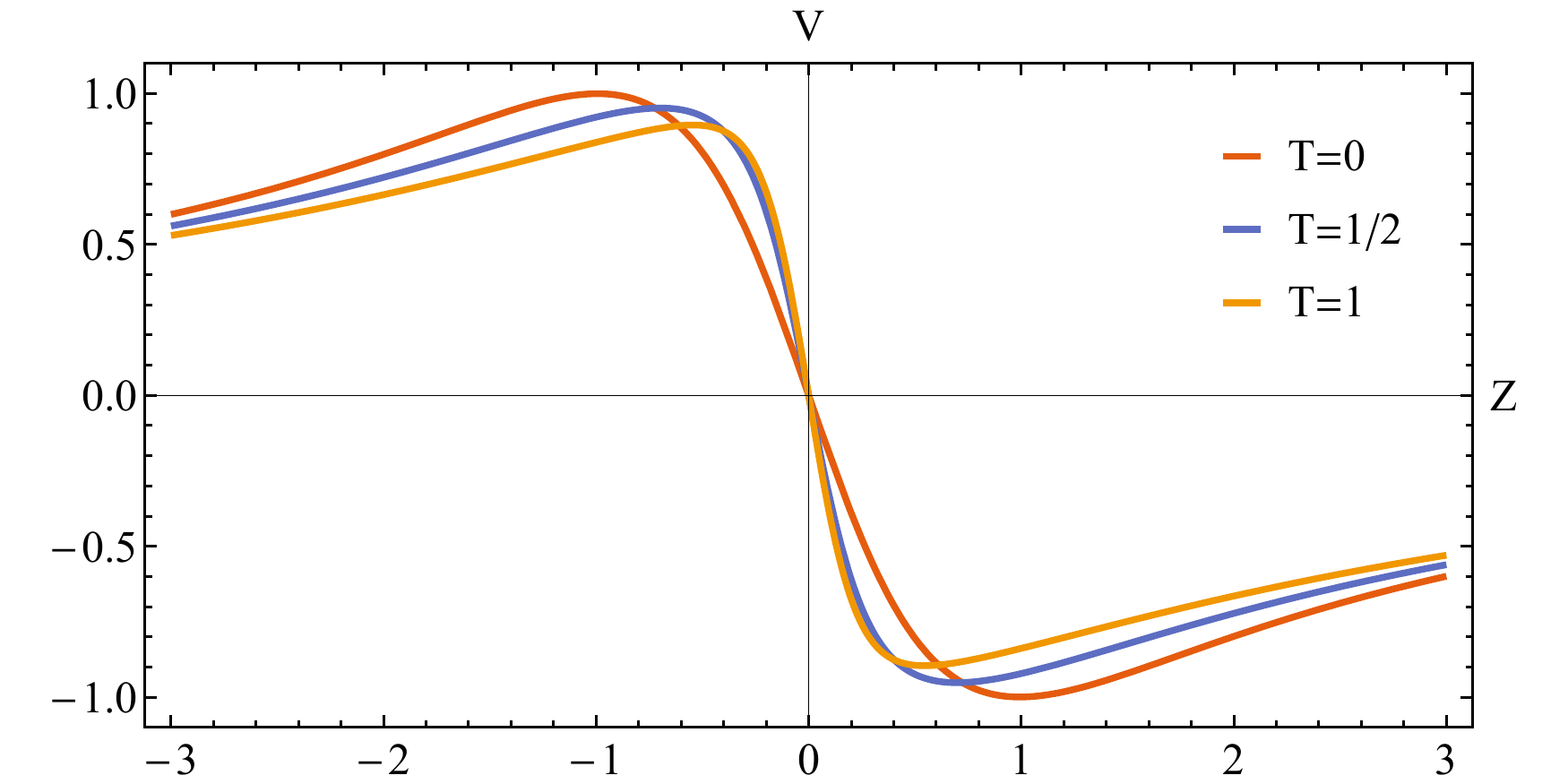}
	\caption{The solution of Burgers equation (\ref{eq:burgers}) for $s=10$}
\end{figure}

\begin{figure}[h] \centering
	\includegraphics[height=0.35\textheight]{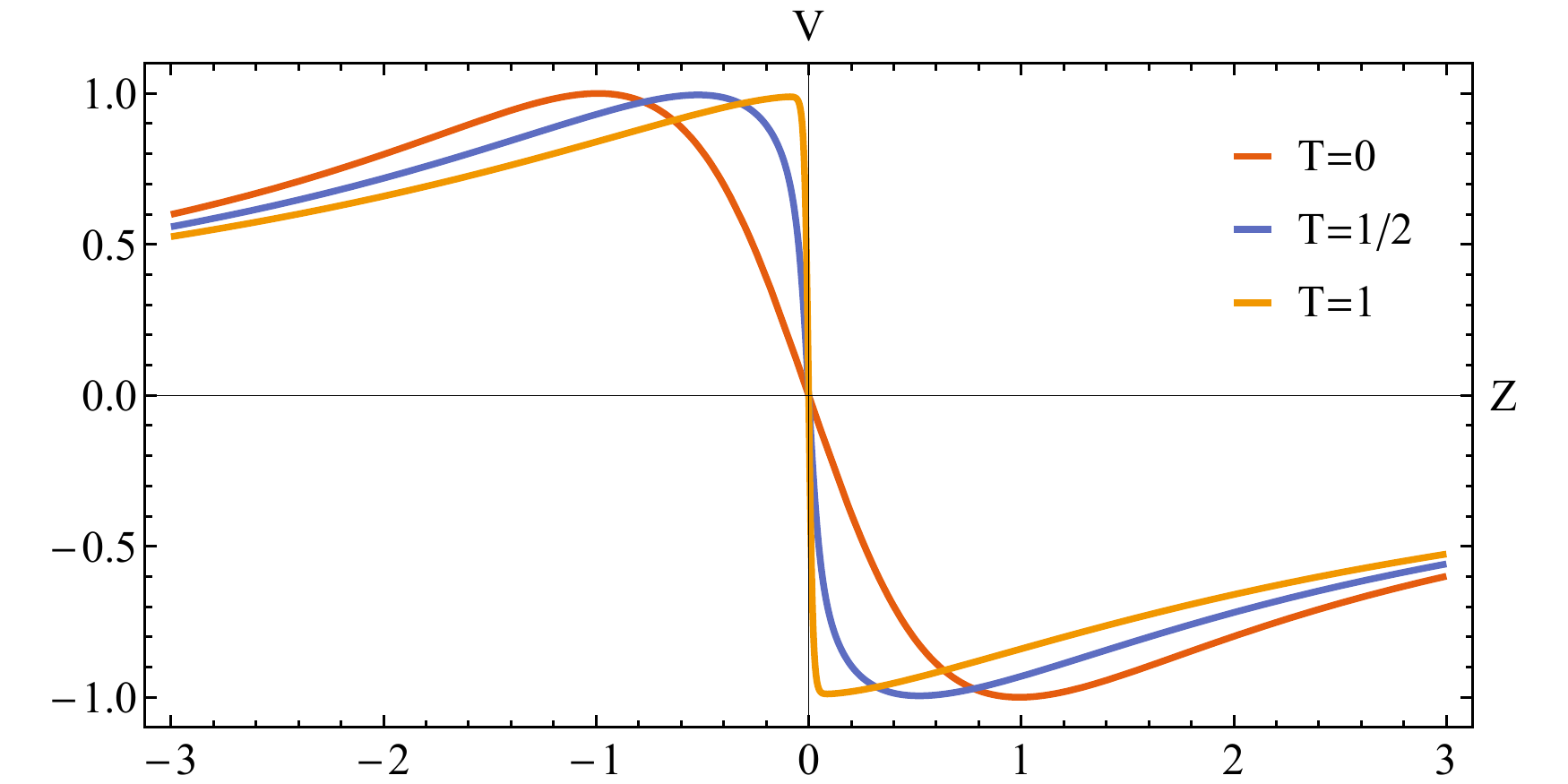}
	\caption{The solution of Burgers equation (\ref{eq:burgers}) for $s=100$}
\end{figure}
To use this solution for verification of numerical methods one can
choose the $Z$ domain as an interval $[-A,A]$ and impose the boundary
conditions using $\mp V(s,T,A)$. This expression is a rational function of $T$ and if $A\gg1$ it has a slowly changing value close to zero.
\section{Case $s=\infty$}
The expression (\ref{eq:otvet}) obtained for a finite $s$
cannot be easily generalized for $s\rightarrow\infty$. We consider this case
separately:
\begin{equation}
\frac{\partial V}{\partial T}+V\frac{\partial V}{\partial Z}=0,\label{eq:invicid}
\end{equation}
\[
V(0,Z)=-\frac{2Z}{1+Z^{2}}.
\]
A well known implicit solution
\begin{equation}
V(T,Z)=f(Z-V(T,Z)T)\label{eq:implicit}
\end{equation}
would satisfy (\ref{eq:invicid}) and the boundary condition
$V(0,Z)=f(Z)$, therefore $f(Z)=-2Z/(1+Z^{2}).$ By substituting it
into (\ref{eq:implicit}) we obtain a polynomial of the third degree in $V$. After introducing for brevity
\begin{equation}
\phi=Z-VT,\label{eq:phi}
\end{equation}
the equation is written as $-\phi^{3}+Z\phi^{2}+\phi(2T-1)+Z=0.$
Consider one of its roots
\[
\phi_{1}(T,Z)=\frac{1}{3}\left(Z-\frac{Z^{2}+6T-3}{(B+3\sqrt{D})^{1/3}}-(B+3\sqrt{D})^{1/3}\right),
\]
\[
B=-Z(9+9T+Z^{2}),\quad D=3Z^{4}-3Z^{2}(T(T-10)-2)-3(2T-1)^{3}.
\]
According to (\ref{eq:phi}) the solution $V$ is 
\begin{equation}
V=Z/T-\phi_{1}/T.\label{eq:resh}
\end{equation}
For $T<1/2$ it satisfies the equation at any $Z$. If $T=1/2$ and
$Z=0$ the expression $D$ under the radical is zero. This corresponds
to the wave tipping, the derivative $\partial V/\partial Z$ jumps
to $-\infty$ (see Figure 4). For $T>1/2$, $D$ has a negative the sign in a certain
neighborhood of $Z=0$, but one can still choose cubic roots properly
so that $\phi_{1}$ will be a real number. One can construct a multivalued
solution for $T>1/2$ by choosing all real roots among $\phi_{1},\phi_{2},\phi_{3}$.
\begin{figure}[h] \centering
	\includegraphics[height=0.35\textheight]{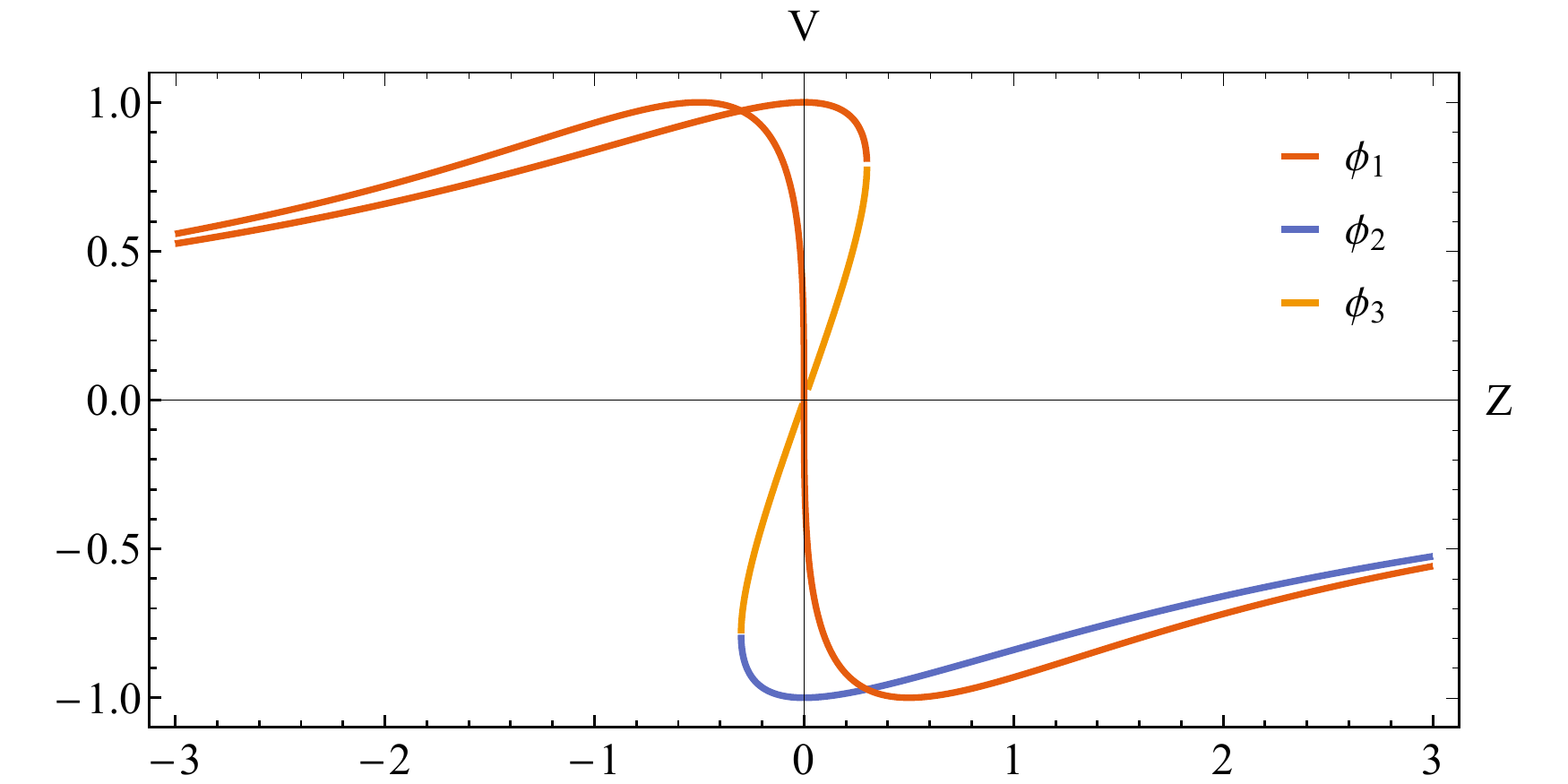}
\caption{Invicid solution for $T=1/2$ and its multi-valued continuation for
	$T=1$.}
\end{figure}
To obtain the single-valued solution we use symmetry considerations.
Namely, the initial condition and equation do not change under transformation
$Z\rightarrow-Z$, $V\rightarrow-V$. Therefore, the solution should
not either: $V(-Z,T)=-V(Z,T).$ The solution obtained with $\phi_{1}$
(orange curve on the Figure 4) is smooth for all $T$ if $Z<0$
and also satisfies the initial condition. By continuing it (anti)symmetrically for $Z>0$
\[
\tilde{V}=\text{sign}(Z)\frac{|Z|+\phi_{1}(T,-|Z|)}{T}
\]
we obtain a discontinuity at $Z=0$ for $T>1/2$. This is a stationary shock ($Z\equiv0$),
and in this case the Rankine-Hugonio condition is
\[
V(+0,T)+V(-0,T)=0.
\]
$\tilde{V}$ satisfies this relation, therefore it is a weak solution of the problem. One can observe that by comparing viscous solutions for large $s$ with $\tilde{V}$ for $T=1$ (see Figure 5). 
\begin{figure}[h] \centering
	\includegraphics[height=0.35\textheight]{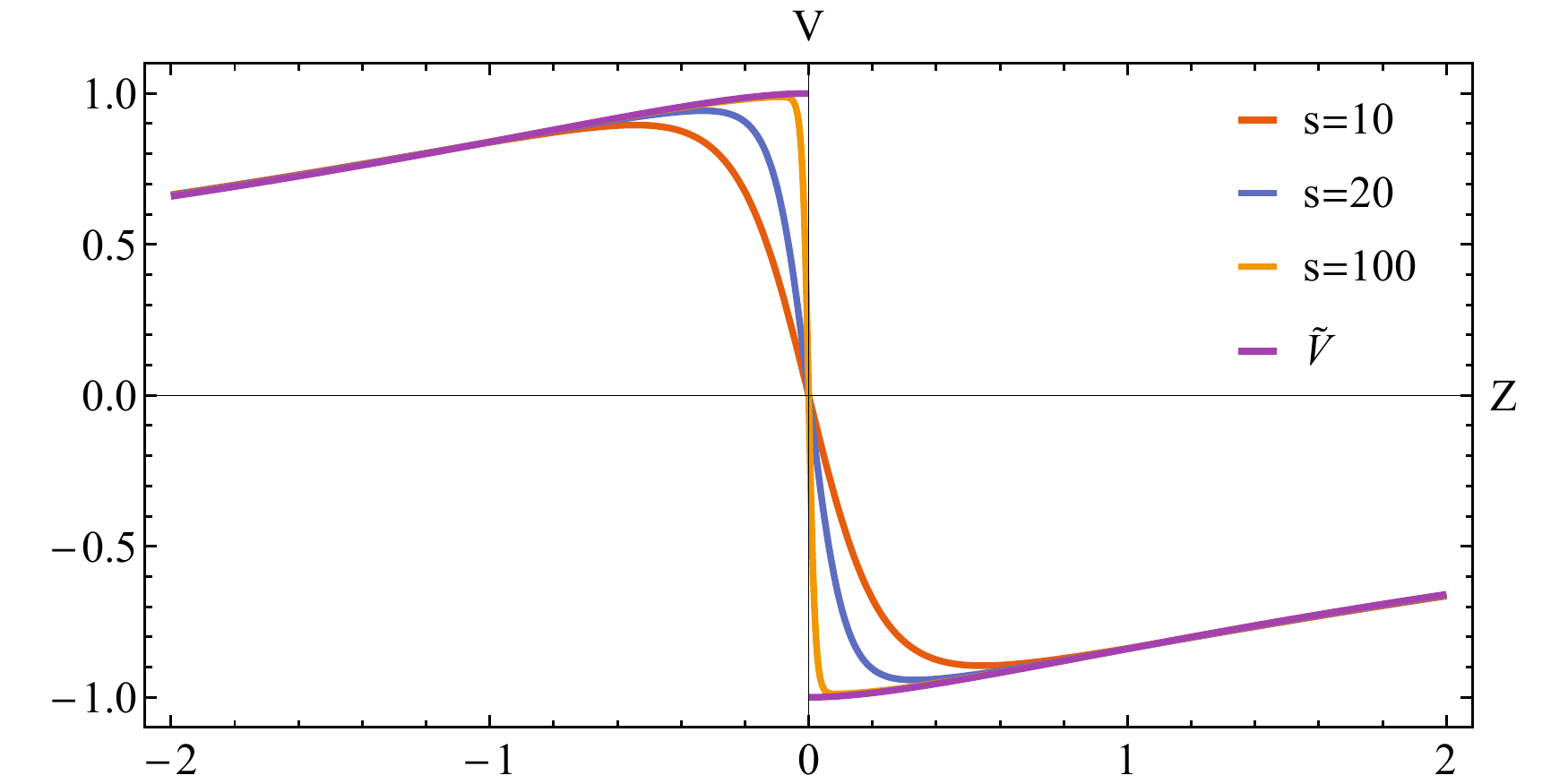}
	\caption{Smooth viscous solutions approaching discontinuous $\tilde{V}$ as the Reynolds number $s$ increases}
\end{figure}
\section{Asymptotic for $s\rightarrow\infty$}
\subsection{Convergence to the invicid solution}
\begin{figure}[h] \centering
	\includegraphics[height=0.4\textheight]{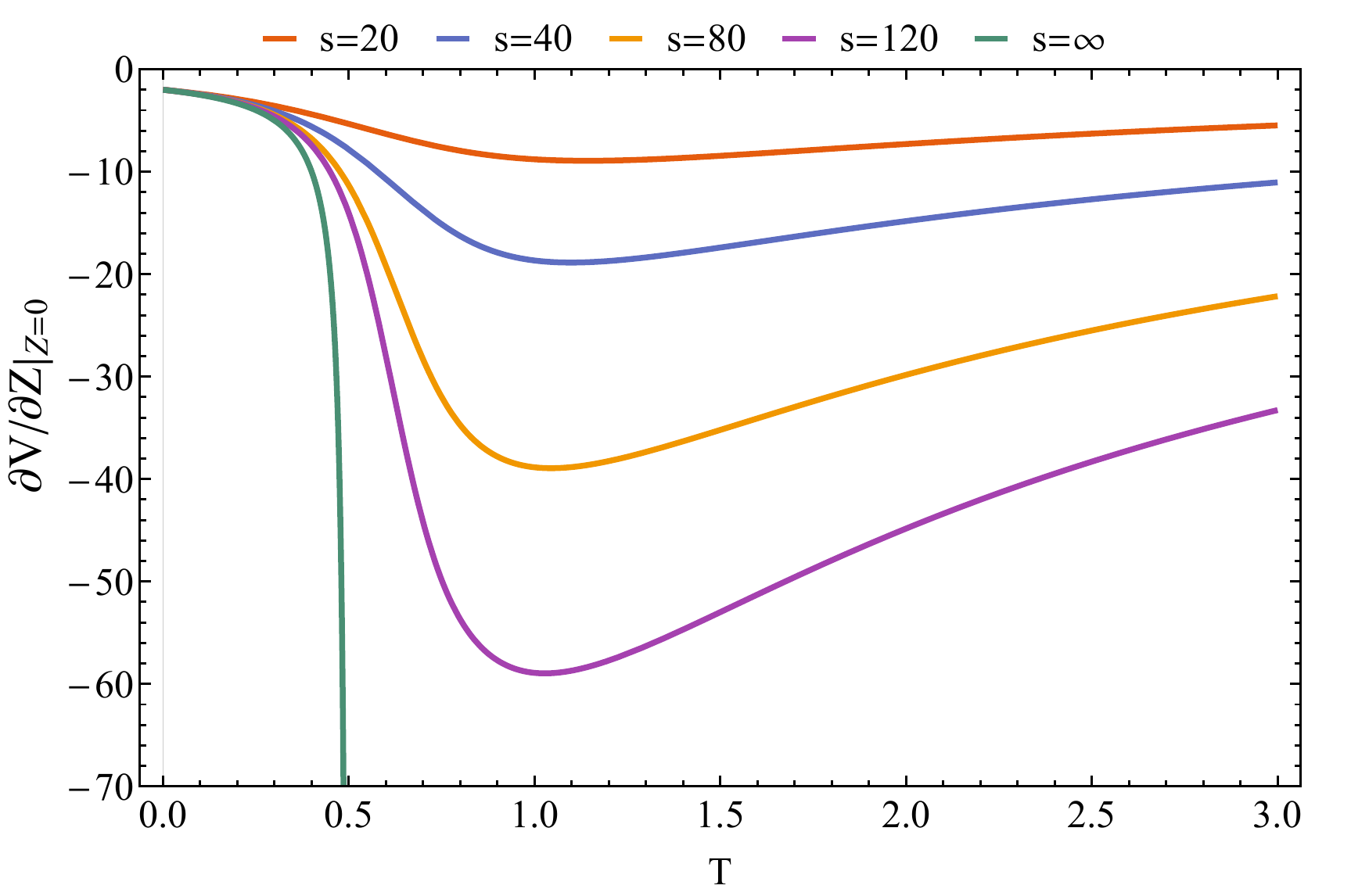}
	\caption{The slope of different solutions at the tipping point $Z=0$ for various
		$s$}
\end{figure}
After looking at Figure 6 one can notice that in contrast with the inviscid case,
in which the derivative jumps to $-\infty$ at the tipping time $T=1/2$ and then the
shock forms, the minimal derivative for large but finite $s$ occurs
at $T\sim1$ and is approximately $-s/2$. Let us inspect this value
more carefully. By differentiating $V$ with respect to $Z$ and putting
$Z=0$ 
\[
\frac{\partial V}{\partial Z}=-\frac{2}{s}\frac{\partial}{\partial Z}\left(\frac{\partial W/\partial Z}{W}\right)\stackrel{Z\rightarrow0}{=}-\frac{2}{s}\left.\frac{\partial^{2}W/\partial Z^{2}}{W}\right|_{Z=0}
\]
($W$ has only even powers of $Z$ therefore $\left.\partial W/\partial Z\right|_{Z=0}=0$).
The denominator is the coefficient of $Z^{0}$ in (\ref{eq:W}),
and the numerator is the doubled coefficient of $Z^{2}$. Luckily,
the sum over $\alpha$ in (\ref{eq:W}) is expressed via Hypergeometric function:
\begin{equation}
\sum_{\alpha=k}^{s/2}\frac{(2\alpha)!}{\alpha!(\frac{s}{2}-\alpha)!(\alpha-k)!}\left(\frac{T}{s}\right)^{\alpha-k}=\frac{\left(s/T\right)^{k+1/2}\frac{s}{2}!}{2\sqrt{\pi}(\frac{s}{2}-k)!}\Gamma(\frac{1}{2}+k)U(\frac{1}{2}+k,\frac{3+s}{2},\frac{s}{4T})\label{eq:coef}
\end{equation}
\[
U(a,c,z)=\frac{1}{\Gamma(a)}\int_{0}^{\infty}e^{-zt}t^{a-1}(1+t)^{c-a-1}dt
\]
After substituting $k=0;1$ into (\ref{eq:coef}), dividing by $2^{k}k!(2k-1)!!$ and simplifying, we obtain
\begin{equation}
\left.\frac{\partial V}{\partial Z}\right|_{Z=0}=-\frac{s}{2T}\frac{U(\frac{3}{2},\frac{3+s}{2},\frac{s}{4T})}{U(\frac{1}{2},\frac{3+s}{2},\frac{s}{4T})}.\label{eq:naklon}
\end{equation}
At $s\rightarrow\infty$ the second and the third arguments of $U$
tend to infinity $c,z\rightarrow\infty$, and there are three different
asymptotic depending on the ratio $z/c$. If $T<1/2$ then $z/c>1$, and the expansion according to \cite{temme2015asymptotic} is
\[
U(a,c,z)\sim(z-c)^{-a}\left(1+\frac{a(1+a)(c-2z)}{2(z-c)^{2}}+O\left(\frac{1}{c^{2}}\right)\right).
\]
By substituting it into (\ref{eq:naklon}) and expanding the
result, we obtain the asymptotic of the viscous slope for $T<1/2$
\begin{equation}\label{expt12}
\left.\frac{\partial V}{\partial Z}\right|_{Z=0}\sim-\frac{2}{1-2T}+\frac{12T}{(1-2T)^{3}}\frac{1}{s}+O\left(\frac{1}{s^{2}}\right).
\end{equation}
For the inviscid solution (\ref{eq:resh}), unsurprisingly
\begin{equation}
\left.\frac{\partial V}{\partial Z}\right|_{Z=0}^{s=\infty}=-\frac{2}{1-2T}.\label{eq:slopeinv}
\end{equation}
For $T<1/2$ at $Z=0$ viscous solutions are equal to the inviscid one, therefore
the discrepancy between their slopes hints on the speed of convergence of the viscous solutions to the inviscid case (at least in a small neighborhood of $Z=0$) as the Reynolds number approaches infinity. However, expression (\ref{expt12}) can not be used near the tipping time. At $T=1/2$ the proper expansion of $U$ at $c,z\rightarrow\infty$ is for which $z/c=1$
\[
U(a,c,c)\sim\sqrt{\pi}(2c)^{-a/2}\left(\frac{1}{\Gamma((a+1)/2)}+O\left(\frac{1}{\sqrt{c}}\right)\right)\text{.}
\]
Plugging it into (\ref{eq:naklon}) and expanding the result yields
\begin{equation}\label{tippingG}
\left.\frac{\partial V}{\partial Z}\right|_{Z=0;T=1/2}\sim-\frac{\Gamma(3/4)}{\Gamma(5/4)}\sqrt{s}+O(1)\simeq-1.35\sqrt{s}+O(1).
\end{equation}
The gradient of the inviscid solution jumps to $-\infty$ and the viscous
solution catches up with a speed proportional to $\sqrt{s}.$
\subsection{Extremum of the gradient}
Figure 3 suggests that the maximum slope occurs at $Z=0$. Indeed,
at this line the equation turns into $\partial^{2}V/\partial Z^{2}=0$,
so any point of it is critical for $\partial V/\partial Z$. To
find the critical time $T_{\text{c}}$ consider the derivative of
(\ref{eq:naklon}) with respect to $T$. It can be evaluated
using the relation
\[
\frac{\partial}{\partial z}U(a,c,z)=-a\cdot U(a+1,c+1,z),
\]
and expanding the results. We expect $T_{\text{c}}>1/2$ therefore $z/c<1$, so the appropriate
asymptotic of $U$ at $c,z\rightarrow\infty$ is
\begin{equation}
U(a,c,z)\sim\frac{\sqrt{2\pi}}{\Gamma(a)}c^{c-3/2}(1-z/c)^{a-1}\exp(z-c)\sum_{p=0}^{\infty}\frac{d_{p}}{c^{p}}\label{eq:expans}
\end{equation}
where $d_{0}=1$ and $d_{p}$ are described in \cite{temme2015asymptotic}. Using three terms, we obtain
\begin{equation}
\frac{\partial}{\partial T}\left.\frac{\partial V}{\partial Z}\right|_{Z=0}\sim\frac{s(-1+T)}{T^{3}}+\frac{1-4T}{T^{2}(2T-1)^{2}}+O\left(\frac{1}{s}\right)=0.\label{eq:extT}
\end{equation}
Solving (\ref{eq:extT}) for $T$ and expanding the result gives
\begin{equation}
T_{\text{c}}\sim1+\frac{3}{s}+O\left(\frac{1}{s^{2}}\right).\label{eq:tmax}
\end{equation}
By substituting (\ref{eq:tmax}) into (\ref{eq:naklon}) and expanding the result we obtain
\[
\left.\frac{\partial V}{\partial Z}\right|_{Z=0;T=T_{\text{c}}}\sim-\frac{s}{2}+1+\frac{7}{2s}+O\left(\frac{1}{s^{2}}\right).
\]
This gradient decreases much faster than the one at the moment
of wave tipping (\ref{tippingG}). To evaluate the Hessian of the slope we obtain similarly:
\[
\frac{\partial}{\partial Z^{2}}\left.\frac{\partial V}{\partial Z}\right|_{Z=0;T=T_{\text{c}}}\sim\frac{s^{3}}{4}-s^{2}-\frac{5s}{2}+O\left(1\right).
\]
The mixed derivative $\partial^{2}/\partial Z\partial T$ of the slope is zero
and $\partial^{2}/\partial T^{2}$ can be calculated directly from (\ref{eq:extT}), and it is positive. Therefore the Hessian is positively defined for large enough $s$ and the point $Z=0$, $T=T_{c}$ is a local minimum.
\section{Final remarks}
We used the fact that the initial condition $V(0,Z)=-\partial/\partial Z\log[1+Z^{2}]=-2Z/(1+Z^{2})$ led to the factor $(1+\widehat{Z}^{2})^{s/2}$ in the integral with $d\widehat{Z}$ (\ref{integral}), and this factor can be expanded using binomial
coefficient. Multinomial coefficients can expand powers
of three and more summands, this allows to obtain physically viable
rational solutions for the initial conditions of a form $V(0,Z)=-\partial/\partial Z\log[\sum_{l=0}^{2n}c_{l}Z^{l}]$, where $c_{l}$ ensure no real roots. In this general case, however, the inviscid solution cannot be expressed in radicals.







\end{document}